\documentclass[
 aip,
 jcp,
 amsmath,
 amssymb,
]{revtex4-1}
\usepackage[utf8]{inputenc}
\usepackage{array}
\usepackage{amssymb}
\usepackage{amsmath}
\usepackage{float}
\usepackage{xcolor}
\usepackage{graphicx}
\usepackage{braket}

\begin{document}

\title{Divergences in classical and quantum linear response and equation of motion formulations}
\author{Erik Rosendahl Kjellgren}
\email{kjellgren@sdu.dk}
\affiliation{Department of Physics, Chemistry and Pharmacy,
University of Southern Denmark, Campusvej 55, 5230 Odense, Denmark.}
\author{Peter Reinholdt}
\affiliation{Department of Physics, Chemistry and Pharmacy,
University of Southern Denmark, Campusvej 55, 5230 Odense, Denmark.}
\author{Karl Michael Ziems}
\affiliation{Department of Chemistry, Technical University of Denmark, Kemitorvet Building 207, DK-2800 Kongens Lyngby, Denmark.}
\author{Stephan P. A. Sauer}
\affiliation{Department of Chemistry, University of Copenhagen, DK-2100 Copenhagen \O.}
\author{Sonia Coriani}
\affiliation{Department of Chemistry, Technical University of Denmark, Kemitorvet Building 207, DK-2800 Kongens Lyngby, Denmark.}
\author{Jacob Kongsted}
\affiliation{Department of Physics, Chemistry and Pharmacy,
University of Southern Denmark, Campusvej 55, 5230 Odense, Denmark.}
\date{\today}

\begin{abstract}
Calculating molecular properties using quantum devices can be done through the quantum linear response (qLR) or, equivalently, the quantum equation of motion (qEOM) formulations.
Different parameterizations of qLR and qEOM are available, namely naive, projected, self-consistent, and state-transfer.
In the naive and projected parameterizations, the metric is not the identity, and we show that it depends on the redundant orbital rotations.
This dependency may lead to divergences in the excitation energies for certain choices of the redundant orbital rotation parameters in an idealized noise-less setting.
Further, this leads to significant variance when calculations include statistical noise from finite quantum sampling.
\end{abstract}


\maketitle

\section{Introduction}

An important feature of quantum chemistry is its ability to predict the properties of a molecular system, such as its UV/Vis or X-ray absorption spectra, i.e., the system's electronic excitation energies and associated absorption strengths.
The calculation of many such molecular response properties can be achieved based on the equation of motion (EOM) \cite{rowe_equations-of-motion_1968} or linear response (LR) \cite{Olsen1985} formulations.
After obtaining a ground-state wave function, response properties (e.g., excitation energies or polarizabilities) can be calculated in the framework of LR or EOM.
In this work, we consider variational wave functions, for which conventional examples are parameterizations such as Hartree-Fock or Kohn-Sham Density Functional Theory (KS-DFT) for weakly correlated systems, 
or Multiconfigurational Self-Consistent Field (MCSCF) methods such as restricted active space (RAS)\cite{Olsen1988,Malmqvist1990} and complete active space (CAS)\cite{Siegbahn1980,Roos1980,Siegbahn1981}
for strongly correlated systems.
One of the main challenges with MCSCF methods is the exponential scaling with respect to the number of active orbitals.
An emerging technology that might solve the scaling issues of MCSCF methods is the use of noisy-intermediate-scale quantum (NISQ) devices.
The predominant approach to exploit these NISQ devices for quantum chemical calculations is currently the variational eigenvalue solver (VQE) approach,\cite{peruzzo_variational_2014, jarrod_theory_2016} which essentially produces expectation values of a Hermitian operator over a state prepared on the quantum device. The wave functions representing these states have to be unitary. Originally, a unitary coupled cluster ansatz (UCC) was employed,\cite{bartlett1989alternative, anand_quantum_2022} but later also orbital-optimized unitary coupled cluster (oo-UCC) wave functions,\cite{sokolov_quantum_2020, mizukami_orbital_2020, bierman_improving_2023} CASSCF type of wave functions\cite{Fitzpatrick2022} or equivalent oo-UCC wave functions with an active space\cite{Jensen2024, Ziems2024} have been proposed. 

A quantum version of the classical EOM approach, dubbed qEOM, has been proposed to obtain electronic excitation energies on quantum computers.\cite{McClean2017, Ollitrault2020} 
Various variants of this approach have been presented, among them the multi-component-EOM,\cite{Pavoevi2021a} QED-EOM,\cite{Pavoevi2021} spin-flip qEOM,\cite{Pavoevi2023}, qEOM for thermal averages of quantum states\cite{https://doi.org/10.48550/arxiv.2406.04475}, a version using self-consistent excitation operators (q-sc-EOM),\cite{Asthana2023} a version of q-sc-EOM using a Davidson algorithm to solve the eigenvalue problem,\cite{Kim2023} and finally a version employing an orbital-optimized active space UCC approach, called the oo-VQE-qEOM method.\cite{Jensen2024}
More recently, a quantum algorithm version of linear response theory (qLR) has also been developed using both self-consistent and projected excitation operators, i.e., a q-sc-LR and a q-proj-LR approach.\cite{Kumar2023} In a later paper, Ziems et al.\cite{Ziems2024} extended qLR to near-term quantum computers by deriving qLR for orbital-optimized active-space wave functions. The authors formulated eight different versions of linear response theory and analyzed their applicability for quantum computer implementations. Very recently, q-naive-LR has also been formulated in terms of reduced density matrices by von Buchwald et al. \cite{Von_Buchwald2024-pp}. Finally, Reinholdt et al.\cite{Reinholdt2024} implemented a version of q-sc-LR, where the elements of the electronic Hessian matrix are never calculated explicitly, i.e., based on a Davidson algorithm, which corresponds to what is normally done in classical quantum chemistry programs like, e.g., DALTON.\cite{dalton}

In some of these approaches, the calculation of electronic excitation energies implies the solution of a generalized eigenvalue problem, i.e., an eigenvalue problem with a metric matrix $\boldsymbol{S}^{[2]}$, which has to be non-singular to obtain solutions. In the present work, we investigate the naive and projected formulations of qLR with a simple model system (the helium atom with a full configuration wave function) and show how the calculation of the excitation energy is affected by the choice of the otherwise redundant orbital rotation coefficient. In addition, we discuss how the introduction of shot noise affects the result of our analysis, which is relevant when the calculation is carried out on a quantum computer simulator.

\section{Theory}

For unitary coupled cluster wave functions, the linear response (LR) and equation of motion (EOM)  formulations are identical.\cite{Taube2006} However, the language used in this work is within the linear response framework. We stress that the conclusions apply equally to the EOM formulations.

\subsection{Unitary Coupled Cluster}

In unitary coupled-cluster theory (UCC), the wave function is constructed as
\begin{equation}
    \left|\text{UCC}\right> = \exp\left(\sum_i\theta_i\left(\hat{T}_i - \hat{T}_i^\dagger\right)\right)\left|\text{CSF}\right> = \exp\left(\sum_i\theta_i\hat{\sigma}_i\right)\left|\text{CSF}\right>
\end{equation}
where CSF is a reference configuration state-function, $\theta_i$ are the cluster-amplitudes and $\hat{T}_i$ the cluster operators (up to doubles) of the form
\begin{align}
    \hat{T}^{(1)}_{pq} &= \frac{1}{\sqrt{2}}\hat{E}_{pq}\\
    \hat{T}^{(2)}_{pqrs} &= \frac{1}{2\sqrt{\left(1+\delta_{pr}\right)\left(1+\delta_{qs}\right)}}\left(\hat{E}_{pq}\hat{E}_{rs} + \hat{E}_{pr}\hat{E}_{sq}\right)\\
    \hat{T}^{\prime(2)}_{pqrs} &=\frac{1}{2\sqrt{3}}\left(\hat{E}_{pq}\hat{E}_{rs} - \hat{E}_{ps}\hat{E}_{rq}\right)
\end{align}
with the singlet single excitation operator, $\hat{E}_{pq} = \hat{a}_{p,\alpha}^\dagger\hat{a}_{q,\alpha} + \hat{a}_{p,\beta}^\dagger\hat{a}_{q,\beta}$.
The presented cluster operators are spin-adapted singlets to guarantee particle and spin preservation.\cite{Paldus1977,Piecuch1989,Packer1996}

The ground-state UCC energy is found by minimizing the energy with respect to the cluster amplitudes:
\begin{equation}
    E_\text{gs} = \min_{\boldsymbol{\theta}}\left<\text{UCC}(\boldsymbol{\theta})\left|\hat{H}\right|\text{UCC}(\boldsymbol{\theta})\right>
\end{equation}
with the molecular Hamiltonian defined as
\begin{equation}
    \hat{H} = \sum_{pq}h_{pq}\hat{E}_{pq} + \frac{1}{2}\sum_{pqrs}g_{pqrs}\hat{e}_{pqrs}~.
\end{equation}
Here, $\hat{e}_{pqrs} = \hat{E}_{pq}\hat{E}_{rs} - \delta_{qr}\hat{E}_{ps}$ is the two-electron singlet excitation operator, 
$h_{pq}$ are the molecular one-electron integrals, and $g_{pqrs}$ are the molecular two-electron integrals. 

The parameterization of the UCC wave function can be extended by considering orbital rotations, 
\begin{equation}
    \left|\text{oo-UCC}\right> = \exp\left(\frac{1}{\sqrt{2}}\sum_{pq}\kappa_{pq}\left(\hat{E}_{pq}-\hat{E}_{qp}\right)\right)\left|\text{UCC}\right> 
    \equiv
\exp\left(\sum_{pq}\kappa_{pq}\hat{\kappa}_{pq}\right)\left|\text{UCC}\right>\label{eq:ooucc-param}
\end{equation}
The orbital rotations can alternatively be formulated as changing the integrals in the Hamiltonian operator instead of working on the state vector\cite{Helgaker2013-xk},
\begin{align}
    h_{pq}\left(\boldsymbol{\kappa}\right) &= \sum_{p'q'} \left[\exp\left(\boldsymbol{\kappa}\right)\right]_{q'q}h_{p'q'}\left[\exp\left(-\boldsymbol{\kappa}\right)\right]_{p'p}\label{eq:int1e_kappa},\\
    g_{pqrs}\left(\boldsymbol{\kappa}\right) &= \sum_{p'q'r's'}\left[\exp\left(\boldsymbol{\kappa}\right)\right]_{s's}\left[\exp\left(\boldsymbol{\kappa}\right)\right]_{q'q}g_{p'q'r's'}\left[\exp\left(-\boldsymbol{\kappa}\right)\right]_{p'p}\left[\exp\left(-\boldsymbol{\kappa}\right)\right]_{r'r}\label{eq:int2e_kappa}.
\end{align}

The ground state energy is then found by minimizing over the full parameter set,

\begin{equation}
    E_\text{gs} = \min_{\boldsymbol{\theta}, \boldsymbol{\kappa}}\left<\text{UCC}(\boldsymbol{\theta})\left|\hat{H}(\boldsymbol{\kappa})\right|\text{UCC}(\boldsymbol{\theta})\right>.
\end{equation}

Using both the UCC and orbital rotation parameterization, some (if not all) of the orbital rotation parameters can be redundant. In this respect, redundant orbital rotations are defined as
\begin{equation}
   E_\text{gs} = \min_{\boldsymbol{\theta}, \boldsymbol{\kappa}}\left<\text{UCC}(\boldsymbol{\theta})\left|\hat{H}(\boldsymbol{\kappa})\right|\text{UCC}(\boldsymbol{\theta})\right> = \left.\min_{\boldsymbol{\theta},\boldsymbol{\kappa}\setminus\kappa_{pq}}\left<\text{UCC}(\boldsymbol{\theta})\left|\hat{H}(\boldsymbol{\kappa})\right|\text{UCC}(\boldsymbol{\theta})\right>\right|_{\kappa_{pq}\in \mathbb{R}}\label{eq:kappa_redundant_condition}.
\end{equation}
An orbital rotation is thus redundant if the ground state energy can be reached for any value of the orbital rotation parameter. 

The parameterization of the orbital rotations is identical to that of the unitary coupled-cluster singles,

\begin{equation}
    \hat{\kappa}_{pq} = \hat{\sigma}^{(1)}_{pq} = \frac{1}{\sqrt{2}}\hat{E}^-_{pq}
\end{equation}
with $\hat{E}^-_{pq} = \hat{E}_{pq} - \hat{E}_{qp}$. This means that for some particular parameterization choices, all orbital rotations will be redundant.
This is the case for the following parameterization:

\begin{align}
    \exp\left(\sum_{pq}\kappa_{pq}\hat{\kappa}_{pq}+\sum_{pq}\theta_{pq}\hat{\sigma}^{(1)}_{pq}\right)\left|\text{CSF}\right> &= \exp\left(\frac{1}{\sqrt{2}}\left(\sum_{pq}\kappa_{pq}\hat{E}^-_{pq}+\sum_{pq}\theta_{pq}\hat{E}^-_{pq}\right)\right)\left|\text{CSF}\right>\\
    &= \exp\left(\frac{1}{\sqrt{2}}\sum_{pq}\left(\kappa_{pq}+\theta_{pq}\right)\hat{E}^-_{pq}\right)\left|\text{CSF}\right>
\end{align}
as $\kappa_{pq}$ covers exactly the same space as $\theta_{pq}$.
However, for the parameterization presented in Eq.\ (\ref{eq:ooucc-param}), the identification of redundant orbital rotations is not trivial since

\begin{equation}
    \exp\left(\sum_{pq}\kappa_{pq}\hat{\kappa}_{pq}\right)\exp\left(\sum_{i}\theta_{i}\hat{\sigma}_{i}\right)\left|\text{CSF}\right> \neq \exp\left(\sum_{pq}\kappa_{pq}\hat{\kappa}_{pq}+\sum_{i}\theta_{i}\hat{\sigma}_{i}\right)\left|\text{CSF}\right>.
\end{equation}
The above turns into an equality only for parameterizations where the commutator between the 
$\sum_{pq}\kappa_{pq}\hat{\kappa}$ and $\sum_{i}\theta_{i}\hat{\sigma}$ operators is zero, i.e. $\left[\sum_{pq}\kappa_{pq}\hat{\kappa}_{pq}, \sum_{i}\theta_{i}\hat{\sigma}_{i}\right] = 0$.
Because of this inequality, all orbital rotations might not be redundant unless the UCC excitation order equals the number of electrons, in which case the UCC parameterization is equal to the FCI parameterization, and all orbital rotations are redundant.

\subsection{Linear Response}

A derivation of the linear response equations will not be given here, but instead, the working equations will be presented.
For a derivation of the linear response equations, see Olsen and 
J{\o}rgensen\cite{Olsen1985}, or, in the context of the unitary coupled cluster with different parameterizations, Ziems et al.\cite{Ziems2024}

For the calculation of excitation energies, the linear  (generalized eigenvalue) response
equation is
\begin{align}
 \left( \boldsymbol{E}^{[2]} - \boldsymbol{\omega} \boldsymbol{S}^{[2]} \right)\boldsymbol{X}   = \boldsymbol{0}\label{eq:lr}~,
\end{align}
where the Hessian and metric matrices are defined as
\begin{align}
 \boldsymbol{E}^{[2]} &= \begin{pmatrix}
    \boldsymbol{A} & \boldsymbol{B}\\
      \boldsymbol{B}^* & \boldsymbol{A}^*           
     \end{pmatrix}, \quad 
     \boldsymbol{S}^{[2]} = \begin{pmatrix}
    \boldsymbol{\Sigma} & \boldsymbol{\Delta}\\
     -\boldsymbol{\Delta} ^* &  -\boldsymbol{\Sigma}^*           
     \end{pmatrix}~. \quad
\end{align}
The submatrices are defined according to 
\begin{align}
    \boldsymbol{A} &= \boldsymbol{A}^\dagger,\quad A_{IJ} =  \left<0\left|\left[\hat{R}_{I}^\dagger,\left[\hat{H},\hat{R}_{J}\right]\right]\right|0\right>\\
    \boldsymbol{B} &= \boldsymbol{B}^\mathrm{T},\quad B_{IJ} =  \left<0\left|\left[\hat{R}_{I}^\dagger,\left[\hat{H},\hat{R}_{J}^\dagger\right]\right]\right|0\right>\\
    \boldsymbol{\Sigma} &= \boldsymbol{\Sigma}^\dagger,\quad \Sigma_{IJ} = \left<0\left|\left[\hat{R}_{I}^\dagger,\hat{R}_{J}\right]\right|0\right>\label{eq:Sigma}\\
    \boldsymbol{\Delta} &= -\boldsymbol{\Delta}^\mathrm{T},\quad \Delta_{IJ} = \left<0\left|\left[\hat{R}_{I}^\dagger,\hat{R}_{J}^\dagger\right]\right|0\right>.
\end{align}
with $\left|0\right>$ being the ground-state reference and the operators $\hat{R}$ defined below. 
In Eq.\ (\ref{eq:lr}), $\boldsymbol{\omega}$ is the matrix of excitation energies, which can be found by solving Eq.\ (\ref{eq:lr}) as a generalized eigenvalue problem.
The base excitation operators for the linear response equation are given as:
\begin{align}
    \hat{G}^{(1)}_{ia} &= \frac{1}{\sqrt{2}}\hat{E}_{ai}\label{eq:G1}\\
    \hat{G}^{(2)}_{ijab} &= \frac{1}{2\sqrt{\left(1+\delta_{ab}\right)\left(1+\delta_{ij}\right)}}\left(\hat{E}_{ai}\hat{E}_{bj} + \hat{E}_{aj}\hat{E}_{bi}\right)\label{eq:G2}\\
    \hat{G}^{\prime(2)}_{ijab} &= \frac{1}{2\sqrt{3}}\left(\hat{E}_{ai}\hat{E}_{bj} - \hat{E}_{aj}\hat{E}_{bi}\right).
\end{align}
Here, the spin-adapted version ensures that only singlet excitations will be calculated.
The indices $i$ and $j$ refer to occupied orbitals in the reference CSF, and $a$ and $b$ refer to virtual orbitals in the reference CSF.

As anticipated in the introduction, different linear response parameterizations have been proposed.
In the context of quantum equation of motion, the naive parameterization was made by Ollitrault et al.\cite{Ollitrault2020} This parameterization has later seen improvements in the form of the projected parameterization,\cite{Fan2021,Liu2022,Kumar2023} the self-consistent parameterization,\cite{Prasad1985,Asthana2023,Kumar2023} and the state-transfer parameterization.\cite{Ziems2024,Olsen1985}
The different parameterizations are defined as
\begin{align}
    &\hat{R}_I^\text{naive} = \hat{G}_I\\
    &\hat{R}_I^\text{proj} = \hat{G}_I\left|0\right>\left<0\right| - \left<0\left|\hat{G}_I\right|0\right>\\
    &\hat{R}_I^\text{sc} = \boldsymbol{U}\hat{G}_I\boldsymbol{U}^\dagger\\
    &\hat{R}_I^\text{st} = \boldsymbol{U}\hat{G}_I\left|\text{CSF}\right>\left<0\right|
\end{align}
with $\boldsymbol{U} = \exp\left(\sum_I\theta_I\hat{\sigma}_I\right)$.
Explicit expressions for the different linear response formulations can be found in Ziems et al. \cite{Ziems2024}
For Eq.\ (\ref{eq:lr}) to be solvable, the metric, $\boldsymbol{S}^{[2]}$, needs to be non-singular.
For all parameterizations considered in this work, it holds that $\boldsymbol{\Delta} = \boldsymbol{0}$, which means that the matrix 
$\boldsymbol{\Sigma}$, Eq.\ (\ref{eq:Sigma}), needs to be invertible.
This condition is guaranteed for the state-transfer and the self-consistent parameterizations, since $\boldsymbol{\Sigma} = \boldsymbol{I}$.
For the naive parameterization, $\boldsymbol{\Sigma}$ takes the form
\begin{equation}
    \Sigma^\text{naive}_{IJ} = \left<0\left|\hat{G}_{I}^\dagger\hat{G}_{J}\right|0\right> - \left<0\left|\hat{G}_{J}^\dagger\hat{G}_{I}\right|0\right>,
\end{equation}
while for the projected parameterization, we obtain:

\begin{equation}
    \Sigma^\text{proj}_{IJ} = \left<0\left|\hat{G}_{I}^\dagger\hat{G}_{J}\right|0\right> - \left<0\left|\hat{G}_{I}^\dagger\right|0\right>\left<0\left|\hat{G}_{J}\right|0\right>.
\end{equation}
These two forms of the metric are not guaranteed to be invertible.
The consequences hereof are addressed and numerically illustrated in the following sections.

\section{Computational Details}

The UCCSD naive-LR and proj-LR calculations were performed using SlowQuant.\cite{slowquant}
The helium atom calculations used 6-31G\cite{Ditchfield1971} basis set.
The hydrogen calculations were based on the use of the  STO-3G\cite{hehre1969a} basis set.
All basis sets are as defined on Basis Set Exchange.\cite{pritchard2019a,feller1996a,schuchardt2007a} The tUPS (tiled Unitary Product State approximation)\cite{Burton2024} ansatz was used for the circuit simulations with simulated statistical noise through SlowQuant with Qiskit-aer\cite{Qiskit} shot noise simulator as a backend.

\section{Results}

\subsection{Helium orbital dependence}

The Helium atom in the basis set 6-31G has been selected as a two electrons in two orbitals model system.
Using the UCCSD parameterization of the wave function gives a parameterization equivalent to the FCI wave function because the truncation order equals the number of electrons.
This parameterization only requires two parameters.
In CI space, the wave function takes the form
\begin{equation}
    \left|\text{FCI}\right> = c_{1100}\left|1100\right> + \frac{c_{1001/0110}}{\sqrt{2}}\left(\left|1001\right> - \left|0110\right>\right) + c_{0011}\left|0011\right>\label{eq:2_2_wf}.
\end{equation}

There are three parameters since the solution is expressed in configuration state functions (CSFs).
As the CI coefficients must satisfy the normalization condition $\sum_i c_i^2 = 1$, the CI representation has only two free parameters.
The occupation number vector uses the convention commonly adopted in quantum chemistry, $\left|1_\alpha 1_\beta 2_\alpha 2_\beta ... N_\alpha N_\beta\right>$, where the number is the orbital index and $N$ is the number of spatial orbitals.
The CI representation of the wave function is recovered from the UCC wave function by applying the parameterization to the reference state
\begin{equation}
    \left|0\right> = \exp\left(\theta_1\hat{\sigma}_1+\theta_2\hat{\sigma}_2\right)\left|\text{CSF}\right> = \left|\text{FCI}\right>
\end{equation}
with $\hat{\sigma}_1$ being the anti-hermitian singles cluster operator and $\hat{\sigma}_2$ being the anti-hermitian doubles cluster operator.
Since the wave function is parameterized to recover the FCI wave function, the orbital rotations fulfill the redundancy condition of Eq.\ (\ref{eq:kappa_redundant_condition}).
Using the redundant orbital rotation parameters, $\kappa_{01}$, the molecular orbital coefficients take the following form,

\begin{equation}
    \boldsymbol{C}\left(\kappa_{01}\right) = \boldsymbol{C}^\text{HF}\exp\begin{pmatrix} 0 & \kappa_{01}\\
    -\kappa_{01} & 0
    \end{pmatrix}.
\end{equation}

Hence, there is only one single orbital rotation parameter.
For this particular model system of two electrons in two orbitals, the Hartree-Fock orbitals are uniquely defined up to a phase.
It should be noted that, in general, the Hartree-Fock solution does not give a unique set of orbitals, i.e. the exact orbitals will depend on the choice of starting guess.

\begin{figure}[H]
    \centering
    \includegraphics[width=0.75\textwidth]{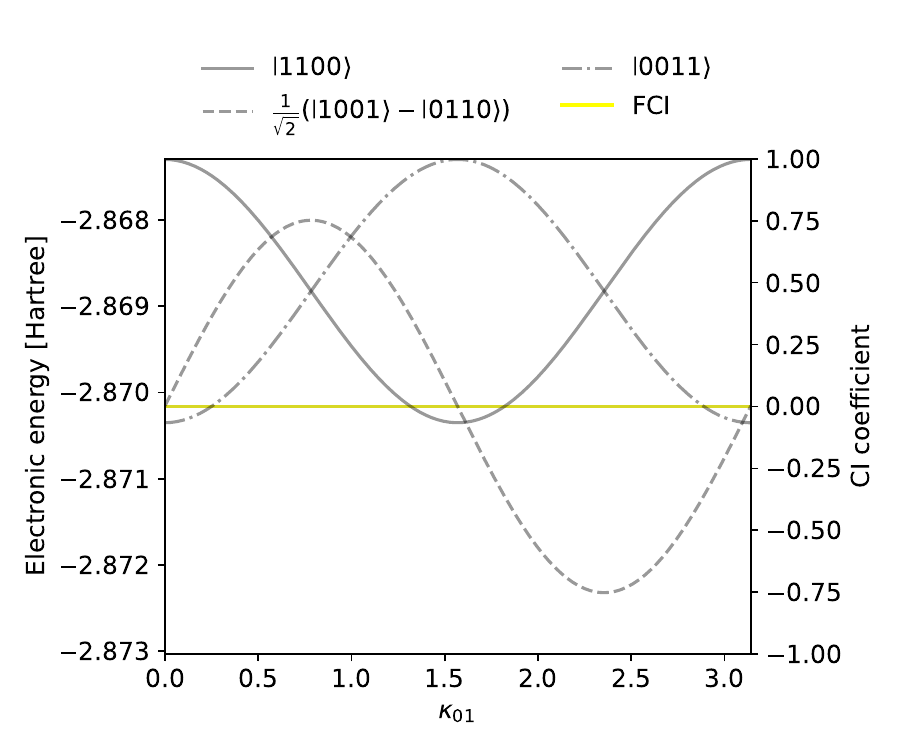}
    \caption{The CI coefficients from the FCI solution as a function of the redundant orbital rotation $\kappa_{01}$. The case of $\kappa_{01}=0$ corresponds to Hartree-Fock orbitals.}
    \label{fig:he_CI}
\end{figure}

In Fig. \ref{fig:he_CI}, the FCI expansion as a function of the redundant orbital rotation parameter, $\kappa_{01}$, is shown.
It can immediately be seen that the specific FCI expansion strongly depends on the orbitals that are used as a reference.
In that context, it is important to point out that Hartree-Fock orbitals are often not the orbitals of choice in multi-configurational calculations, where schemes such as starting from the natural orbitals of relatively inexpensive correlation methods such as MP2,\cite{Mller1934,Jensen1988} or projection schemes to localize orbitals such as AVAS,\cite{Sayfutyarova2017} amongst many other methods, are commonly used.

\subsection{Divergences in naive-LR}

For the two-electrons-in-two-orbitals system, the excitation operators Eq.\ (\ref{eq:G1}) and (\ref{eq:G2}) take the following form
\begin{equation}
    \hat{G}_1 = \frac{1}{\sqrt{2}}\left(\hat{a}^\dagger_{1,\alpha}\hat{a}_{0,\alpha}+\hat{a}^\dagger_{1,\beta}\hat{a}_{0,\beta}\right)\label{eq:G1_22}
\end{equation}
and
\begin{equation}
    \hat{G}_2 = \hat{a}^\dagger_{1,\alpha}\hat{a}_{0,\alpha}\hat{a}^\dagger_{1,\beta}\hat{a}_{0,\beta}\label{eq:G2_22}~.
\end{equation}
This leads to the following metric
\begin{equation}
    \boldsymbol{\Sigma}^\text{naive} = \begin{pmatrix}
        c^2_{1100} - c^2_{0011} & \left(c_{1100} - c_{0011}\right)c_{1001/0110}\\
        \left(c_{1100} - c_{0011}\right)c_{1001/0110} & c^2_{1100} - c^2_{0011}
    \end{pmatrix}
\end{equation}

For the metric to always be invertible, its determinant 

\begin{equation}
    \det(\boldsymbol{\Sigma}^\text{naive}) = \left(c^2_{1100} - c^2_{0011}\right)^2 - \left(\left(c_{1100} - c_{0011}\right)c_{1001/0110}\right)^2\label{eq:naive_det}
\end{equation}
must always be different from zero.
In the limit of $\left|c_{1100}\right| = \left|c_{0011}\right|$, Eq.\ \eqref{eq:naive_det} becomes zero, thus creating a divergence in the naive linear response equations as the metric  $\boldsymbol{S}^{[2]}$ becomes singular.

\begin{figure}[H]
    \centering
    \includegraphics[width=0.75\textwidth]{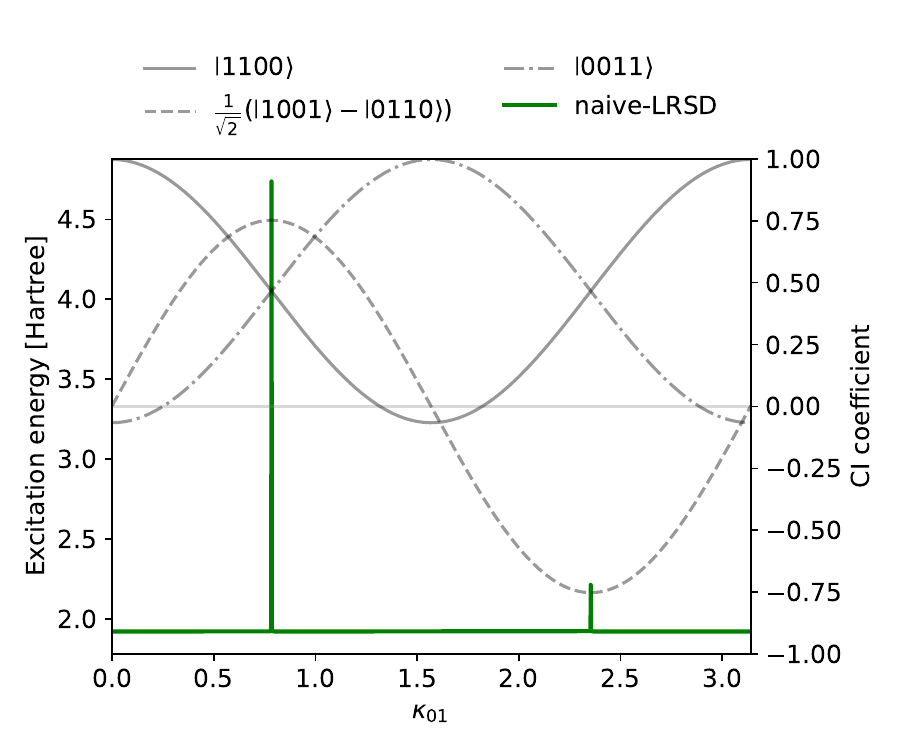}
    \caption{The single excitation energy as a function of the redundant orbital rotation parameter $\kappa_{01}$. $\kappa_{01}=0$ is the Hartree-Fock solution.}
    \label{fig:he_naive}
\end{figure}

In Fig. \ref{fig:he_naive}, we test this numerically by showing the lowest singlet excitation energy of the Helium atom in the 6-31G basis set as a function of the redundant orbital rotation, $\kappa_{01}$.
The excitation energy is constant except for two spikes that appear exactly at the point of $\left|c_{1100}\right| = \left|c_{0011}\right|$.

From the investigations on the Helium atom, it is natural to consider whether Hartree-Fock orbitals would be a good choice to avoid a divergence.
However, Hartree-Fock orbitals do not guarantee that a divergence can be avoided.
This is shown by considering stretched H$_2$ in a minimal basis, where the FCI solution takes the form
\begin{equation}
    \lim_{R_\text{HH}\rightarrow \infty}\left|\text{FCI}\left(\kappa^\text{HF}\right)\right> = \frac{1}{\sqrt{2}}\left|1100\right> - \frac{1}{\sqrt{2}}\left|0011\right>~.
\end{equation}
I.e., $\left|c_{1100}\right| = \left|c_{0011}\right| = 2^{-1/2}$, which will make the metric singular as seen in Eq.\ (\ref{eq:naive_det}).

\begin{figure}[H]
    \centering
    \includegraphics[width=0.75\textwidth]{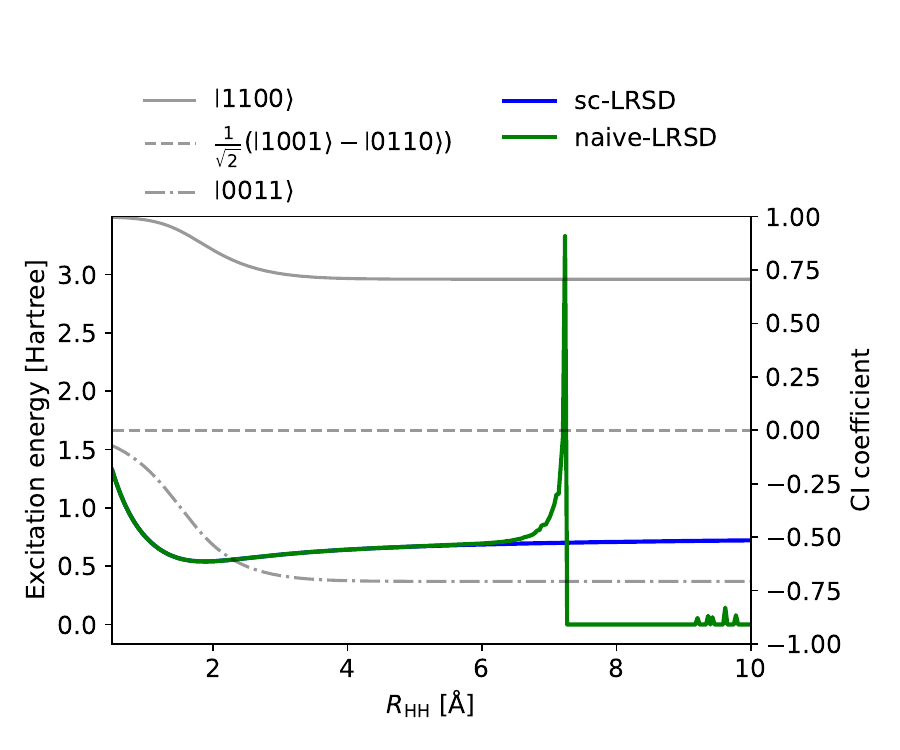}
    \caption{The single excitation energy as a function of the distance between the two hydrogen atoms. $\kappa_{01}=0$ always, i.e. the Hartree-Fock orbitals.}
    \label{fig:h2_naive}
\end{figure}

In Fig. \ref{fig:h2_naive}, the FCI expansion from Hartree-Fock orbitals and the lowest singlet excitation energy calculated with the naive parameterization of the linear response equations can be seen as a function of the distance between the two hydrogen atoms.
With naive-LRSD, the excitation energy diverges as the distance between the hydrogen atoms increases.
The numerical breakdown of the response equations starts to happen already at about 6 Å. As a comparison, the excitation energy calculated using the self-consistent operators is presented in blue and shows no divergence as $\boldsymbol{\Sigma}=\boldsymbol{I}$.

\subsection{Divergences in proj-LR}

Inspecting the projected parameterization and using the same base operators as for the naive parameterization, Eq.\ (\ref{eq:G1_22}) and (\ref{eq:G2_22}), the metric in proj-LR takes the following form,
\begin{equation}
    \boldsymbol{\Sigma}^\text{proj} = \begin{pmatrix}
         c_{1100}^2 + c_{1001/0110}^2 - \left(c_\text{1100}c_\text{1001/0110} + c_\text{0011}c_\text{1001/0110}\right)^2 & \Sigma_{01}\\
         \Sigma_{01} & c_\text{1100}^2-c_\text{1100}^2c_\text{0011}^2
    \end{pmatrix}
\end{equation}
with $\Sigma_{01}=c_{1100}c_{1001/0110} - c_{1100}c_{0011}\left(c_{1100}c_{1001/0110} + c_{0011}c_{1001/0110}\right)$.
This yields the metric determinant, 
\begin{align}
    \det(\boldsymbol{\Sigma}^\text{proj}) &= \left(c_{1100}^2 + c_{1001/0110}^2 - \left(c_\text{1100}c_\text{1001/0110} + c_\text{0011}c_\text{1001/0110}\right)^2\right)\left(c_\text{1100}^2-c_\text{1100}^2c_\text{0011}^2\right)\\\nonumber
    &\quad- \left(c_{1100}c_{1001/0110} - c_{1100}c_{0011}\left(c_{1100}c_{1001/0110} + c_{0011}c_{1001/0110}\right)\right)^2,
\end{align}
which becomes zero at $c_{1100}=0$.
\begin{figure}[H]
    \centering
    \includegraphics[width=0.75\textwidth]{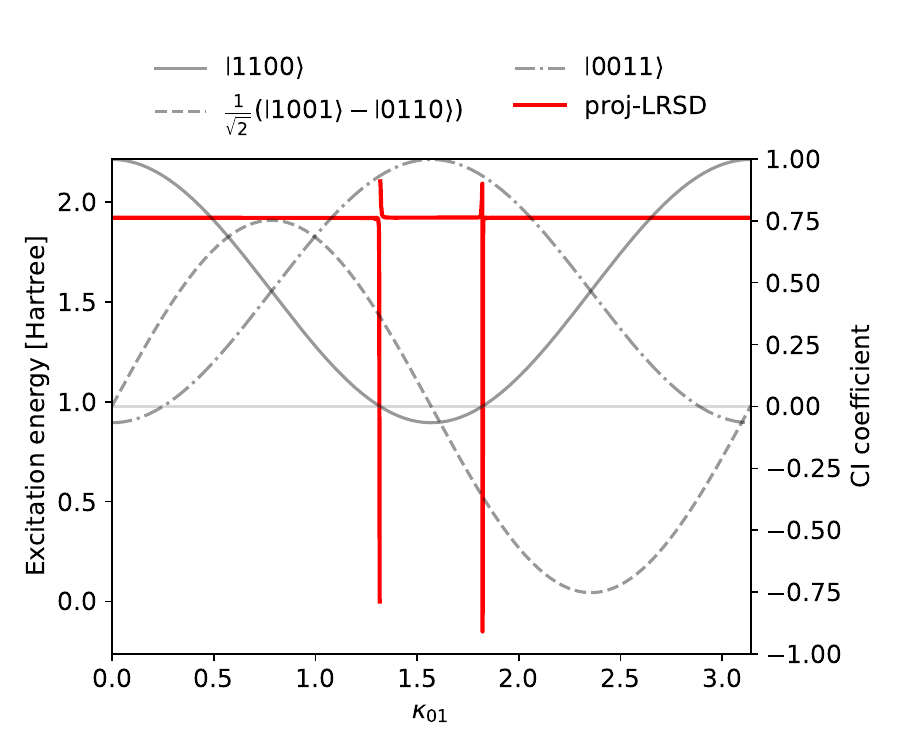} 
    \caption{The single excitation energy as a function of the redundant orbital rotation parameter $\kappa_{01}$. $\kappa_{01}=0$ is the Hartree-Fock solution.}
    \label{fig:he_proj}
\end{figure}

We confirm this numerically in Fig. \ref{fig:he_proj} by calculating the lowest singlet excitation energy using the projected linear response parameterization as a function of the redundant orbital rotation parameter, $\kappa_{01}$.
It can be seen that the excitation energy is constant, except for the two points where $c_{1100}=0$.

\subsection{Effect of noise}
The numerically computed divergences in Fig. \ref{fig:he_naive} and \ref{fig:he_proj} manifest as $\delta$-distributions.
In addition to the numerical instabilities in the response equations, solving the equations using a quantum device will inherently introduce statistical/sampling noise, which will affect the results and the analyses.
The additional effect of sampling noise will be investigated in this section.
On a quantum device (and on circuit simulators), the trotterized version of UCCSD is usually employed. However, this ansatz is not equal to the UCCSD ansatz, and it is not guaranteed that the orbital rotation $\kappa_{01}$ is exactly redundant, i.e., Eq.\ (\ref{eq:kappa_redundant_condition}) might not be fulfilled.
Therefore, the tUPS ansatz with a single layer is used instead because it is flexible enough to solve the two-electrons-in-two-orbitals problem exactly.

\begin{figure}[H]
    \centering
    \includegraphics[width=0.75\textwidth]{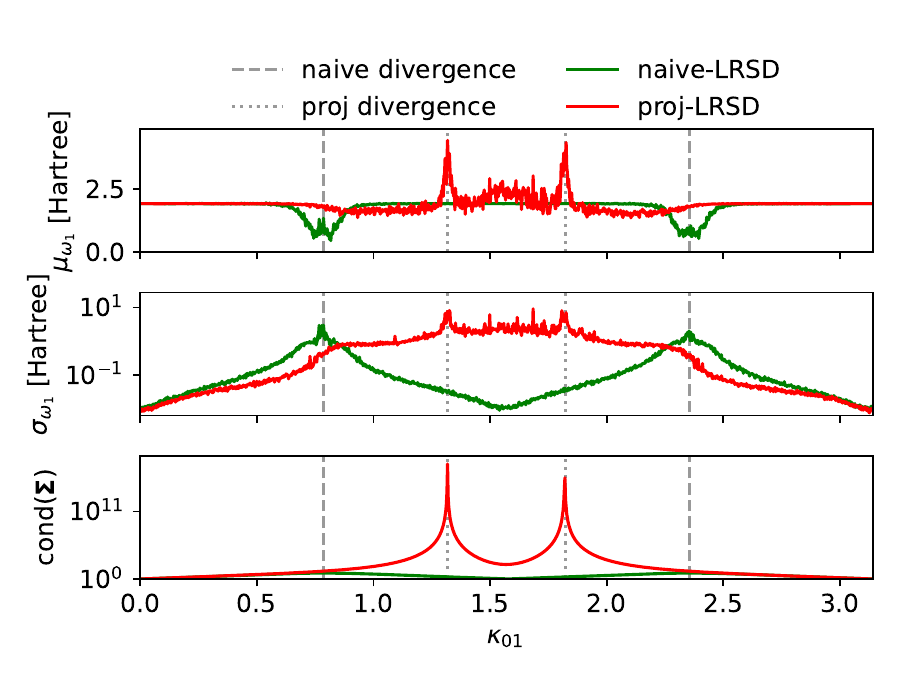} 
    \caption{The mean value and standard deviation of the first excitation energy based on 100 calculations each using 1000 shots. $\mathrm{cond}\left(\boldsymbol{\Sigma}\right)$ is the condition number of $\boldsymbol{\Sigma}$ for the noiseless matrix, calculated using the $L^2$ norm. All quantities are reported as a function of the redundant orbital rotation $\kappa_{01}$.}
    \label{fig:he_noise}
\end{figure}

In Fig. \ref{fig:he_noise}, the mean value and standard deviation of the lowest excitation energy can be seen for 100 samples calculated at each point using shot noise with 1000 shots.
The condition number of the noiseless metric is presented in the bottom panel and can be seen as a measurement of how resilient the metric is to noise.
The figure shows that the calculated excitation energies become unstable in the vicinity of the divergences.
This makes the choice of orbitals a more practical concern when performing noisy calculations.
The standard deviation of naive-LRSD shows a large sensitivity to the orbitals.
Already at $\kappa_{01}\approx0.5$, the standard deviation has increased by an order of magnitude compared to the lowest standard deviation.

A measurement for the stability of the linear response equations is the condition number of the metric, as can be seen in the third panel in Fig. \ref{fig:he_noise}.
For the projected parameterization, the condition number increases rapidly in the region where the excitation energies diverge.
The same huge increase in the condition number cannot be seen for the naive parameterization. This is because, for the two-electrons-in-two-orbitals system, the $\boldsymbol{\Sigma}^\text{naive}$ matrix happens to be a bisymmetric matrix.
The eigenvalues of a two-by-two bisymmetric matrix fulfill $|\lambda_1| = |\lambda_2|$, which in turn means the condition number is very stable.
This will not be the general case for the naive parameterization.

To showcase that this problem is not unique to Helium, we next consider H$_4$ in STO-3G.
Maximizing the condition number with respect to the redundant orbital rotations can be used to find orbitals that will give unstable excitation energies from solving the linear response equations.

\begin{equation}
    \boldsymbol{\kappa}^\text{div} = \arg\max_{\boldsymbol{\kappa}^\text{red}}\left\{\text{cond}\left(\boldsymbol{\Sigma}\left(\boldsymbol{\kappa}^\text{red}\right)\right)\right\}
    \label{eq:bad_orbs}
\end{equation}

This maximization is made while keeping the energy constant by changing $\boldsymbol{\theta}$ accordingly.
For H$_4$ in STO-3G, there are six redundant orbital rotations.

\begin{figure}[H]
    \centering
    \includegraphics[width=0.75\textwidth]{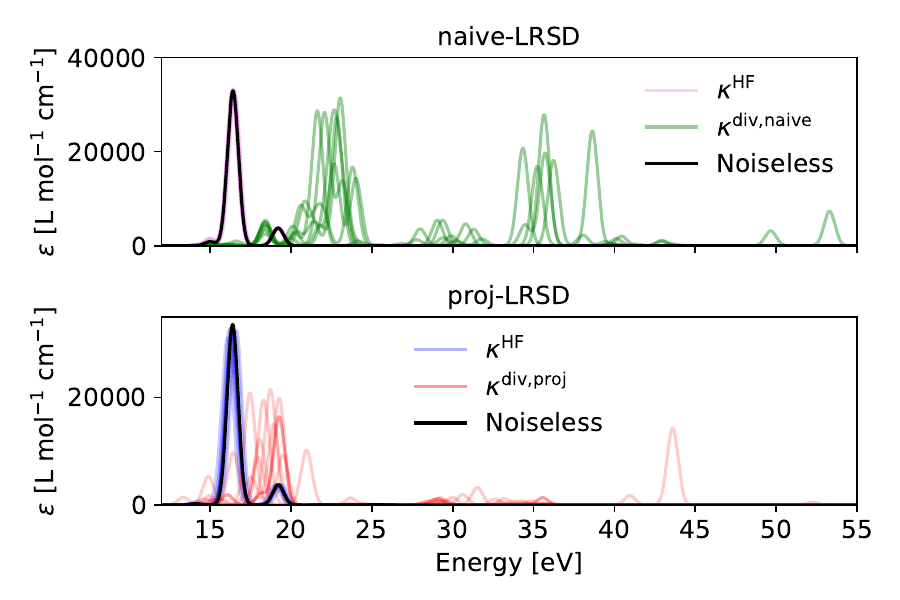} 
    \caption{Calculated absorption spectra of H$_4$ using 100k shots on shot noise simulator.
    Each spectrum was calculated 10 times.
    The H$_4$ molecule is rectangular with the sides being 1.5 Å and 1.8 Å.}
    \label{fig:h4}
\end{figure}

In Fig. \ref{fig:h4}, calculated absorption spectra with simulated shot noise using 100k shots can be seen for H$_4$ using both naive-LRSD and proj-LRSD.
The wave function is parameterized using pp-tUPS with three layers to get a description that is flexible enough to reach the FCI wave function.
$\boldsymbol{\kappa}^\text{div}$ for both naive and projected was found in accordance with Eq. (\ref{eq:bad_orbs}) using the COBYLA\cite{Powell1994} optimizer through SciPy.\cite{2020SciPy-NMeth}
The noiseless condition numbers found are $\mathrm{cond}\left(\boldsymbol{\Sigma}^\text{naive}\left(\boldsymbol{\kappa}^\text{div,naive}\right)\right) = 1.4\times10^6$ and $\mathrm{cond}\left(\boldsymbol{\Sigma}^\text{proj}\left(\boldsymbol{\kappa}^\text{div,proj}\right)\right) = 2.8\times10^{12}$.
This shows that divergences can also occur for larger systems.
It should also be noted that the condition number for the naive metric blows up, which was not the case for the Helium atom in the 6-31G basis set.
In comparison for this system, the Hartree-Fock orbitals give a much lower condition number of the metrics, $\mathrm{cond}\left(\boldsymbol{\Sigma}^\text{naive}\left(\boldsymbol{\kappa}^\text{HF}\right)\right) = 1.5$ and $\mathrm{cond}\left(\boldsymbol{\Sigma}^\text{proj}\left(\boldsymbol{\kappa}^\text{HF}\right)\right) = 1.4$.
The effect of the invertibility of the metric can clearly be seen in Fig. \ref{fig:h4}, where the shot noise calculated spectra using Hartee-Fock orbitals are very close to the noiseless result.
In contrast, the orbitals that are optimized to increase the condition number of the metric give very unreliable spectra.

\section{Conclusion}

In this work, we have shown that divergence may occur in linear response parameterizations based on naive or projected excitation operators.
These divergences are found by changing the redundant orbital rotation parameters, i.e., the ground-state electronic energy is unaffected by these parameters. The divergences appear because the metric in the generalized eigenvalue problem becomes singular for certain values of the redundant orbital rotation parameters within the naive and projected parameterizations. In the noiseless case, this leads to $\delta$-distribution-like divergences in the calculated excitation energies. 
When sampling/statistical noise is included, we show that the variance of the calculated excitation energies significantly increases in a wide neighbourhood around the divergences. Thus, the specific choice of orbitals becomes very important when performing quantum linear response or quantum equation of motion calculations based on the naive or the projected parameterization on circuit simulators or quantum devices. This is caused by the additional variance that different orbitals can introduce to the calculated excitation energies. Quantum linear response and quantum equation of motion based on the self-consistent or state-transfer parameterization will not suffer from the metric being orbital dependent since, in this case, the metric is the identity matrix.

\acknowledgments
We acknowledge the support of the Novo Nordisk Foundation (NNF) for the focused research project ``Hybrid Quantum Chemistry on Hybrid Quantum Computers'' (grant number  NNFSA220080996).

\section*{DATA AVAILABILITY}
The data that support the findings of this study are available from the corresponding author upon reasonable request.

\newpage
\bibliographystyle{unsrt}
\bibliography{literature}

\end{document}